\documentclass[reqno]{amsart}
\usepackage{amssymb}
\usepackage{amssymb,epsfig,graphics}
\usepackage{amsmath}
\usepackage{amscd}
\usepackage{verbatim}
\usepackage[figureleft]{rotating}
\input xypic

\allowdisplaybreaks

\theoremstyle{definition}

\newtheorem{example}{Example}

\def\R{\mathbb R}
\def\Z{\mathbb Z}
\def\N{\mathbb N}

\def\r{\rangle}
\def\l{\langle}

\def\o{\omega}
\def\l{\langle}
\def\r{\rangle}

\begin{document}

\title[Discretization of orbit functions of $G_2$]
{Four types of special functions of  $G_2$\\
and their discretization}

\author{Marzena Szajewska}
\address{Institute of Mathematics, University of Bialystok, Akademicka 2, PL-15-267, Bialystok, Poland.}
\email{marzena@math.uwb.edu.pl}

\date{\today}
 \begin{abstract}\

Properties of four infinite families of special functions of two real variables, based on the compact simple Lie group $G_2$, are compared and described. Two of the four families (called here $C$- and $S$-functions) are well known. New results of the paper are in description of two new families of $G_2$-functions not found in the literature. They are denoted as $S^L$- and $S^S$-functions.

It is shown that all four families have analogous useful properties. In particular, they are orthogonal when integrated over a finite region $F$ of the Euclidean space. They are also orthogonal as discrete functions when their values are sampled at the lattice points $F_M\subset F$ and added up with appropriate weight function. The weight functions are determined for the new families. Products of ten types among the four families of functions, namely $CC$, $CS$, $SS$, $SS^L$, $CS^S$, $SS^L$, $SS^S$, $S^SS^S$, $S^LS^S$ and $S^LS^L$, are completely decomposable into the finite sum of the functions belonging to just one of the families. Uncommon arithmetic properties of the functions are pointed out and questions about numerous other properties are brought forward.
 \end{abstract}

\maketitle
\section{Introduction}\label{Intro}
Discretization of characters of irreducible finite dimensional representations of any compact simple Lie group $G$ was introduced in general in \cite{MP87}. At the time it was motivated by the need to solve large computational problems in Lie theory. A quarter of century later the motivation would rightfully be called processing of digital data from lattices of any dimension, density, and symmetry. Large computations made it a practical imperative to replace the basis of irreducible characters by the basis of constituents of the characters that are symmetric with respect to the Weyl group denoted here $C$-functions. Unlike the characters, $C$-functions are firmed as sums of limited number of exponential terms.

It is perhaps curious to note that such a discretization was made possible only after publication of three papers during the few years preceding  \cite{MP87}. They were the following: (i) a concise description  all conjugacy classes of elements of finite order in any $G$ in \cite{Kac}, then (ii) after it became possible to decompose even the extremely large characters into the sum of the sum of constituent $C$-functions \cite{BMP,MP82}, and finally (iii) after the computation of character values of elements of finite order in $G$ became practical \cite{MP84}. The discretization of the skew-symmetric $S$-functions in \cite{MP06} follows an analogous path.

The $C$- and $S$-functions appear in the Weyl character formula \cite{BS}. The functions have been known, for more than half a century, as different constituent of the character \cite{Bourbaki}. They are well known for the uniformity of their properties across the compact simple Lie groups of any type, in particular their continuous orthogonality when integrated over a finite region $F$ of the real Euclidean space, and their discrete orthogonality when summed up over a lattice fragment in $F$ of any density \cite{MP87,MP06}. In the specific case of $G_2$, they were described in \cite{PZ2} and \cite{PZ3} respectively.

The aim of this paper is to confirm for the simple Lie group $G_2$, the existence and properties of two additional families of special functions denoted $S^L$ and $S^S$. Existence of such functions was first noticed\footnote{In \cite{MotP} different notations are used. Namely $C=C^+$, $S^L=C^-$, $S=S^-$ and $S^S=S^+$.} in  \cite{MotP} for the group $C_2$.
All four families of functions are obtained by a summation of exponential terms over the Weyl group $W(G_2)$ of $G_2$. They are refereed to as orbit functions. In 1-dimension, when the underlying group is the rank one simple Lie group $A_1$ (equivalently $SU(2)$), the $C$- and $S$-functions become the common cosine and sine functions, while $S^L$- and $S^S$-functions have no analog there. We point out in the paper that the families of $S^L$- and $S^S$-functions do not exist for the group $A_2$ (equivalently $SU(3)$).

The history of the functions of the families $S^L$ and $S^S$ is short. It was noted in \cite{MotP} that the 2-variable antisymmetric cosine and symmetric sine can be obtained by summation of appropriate exponential functions over the Weyl group $W(C_2)$ of the simple Lie group $C_2$. Since \cite{MotP} is a special case of \cite{KP4}, where the $n$-variable symmetric and antisymmetric sine and cosine were defined, the families of $S^L$- and $S^S$-functions could have been noted already in \cite{KP4} in the context of the simple Lie groups $C_n$ of any rank.

 The present paper is thus the first to describe the families $S^L$ and $S^S$ of orbit functions outside of the symmetric and antisymmetric generalizations of trigonometric functions, their symmetries are more complicated. Most recently it was shown that the four families of orbit functions exist for all compact simple Lie groups with two root lengths \cite{MMP}.

The new results of the paper, which are the properties of $S^L$- and $S^S$-functions of $G_2$, are brought together and compared with the properties of the $G_2$-orbit functions of all four families.

The second goal of the paper is to describe the properties of the new families of orbit functions when they are discretized, that is when their values are sampled on a lattice fragment $F_M\subset F\subset\R^2$. It turns out that the functions within families $C$, $S$, $S^L$ and $S^S$ are orthogonal when summed up over the same finite fragment $F_M$ of the lattice in $F$ of any density specified by $M\in\N$. Although the same grid $F_M$ may be used for each family, the appropriate weight functions are different. That makes the functions of either family suitable for expansion of 2-dimensional digital data. This has been known for over two decades  \cite{MP87} and used in some very large computations, for example in \cite{GP}. It is also known for $S$-functions \cite{MP06,PZ3}. For  $S^L$ and $S^S$ types. It is exposed only here and in the forthcoming papers \cite{MMP,MotP}.

The definition of the orbit function is contained in Sec.~\ref{OF} together with description of their behavior at the boundary of the domain $F$ and their continuous orthogonality in $F$. It is shown that all the orbit functions are either real or purely imaginary. A simple renormalization makes the later functions also real.

In section \ref{Discret}, orbit functions are discretized. Discrete orthogonality of the orbit functions is proven.

In section \ref{Decomp}, the decomposition of the products of pairs of orbit functions into their sums is described. All 10 products, $CC$, $CS$, $CS^L$, $CS^S$, $SS$, $SS^L$, $SS^S$, $S^LS^L$, $S^LS^S$ and $S^SS^S$, decompose into the sum of orbit functions from just one family. A product of a pair of orbit functions contains either 144 or 72 or 36 exponential terms which can be rewritten as half as many trigonometric function terms. The information is then used to find the recurrence relations for the orbit functions.

Arithmetic properties of the orbit functions remain mostly unexplored in the literature. In section 6, we present just the properties of $C$-functions, directly deduced from the corresponding properties of the characters \cite{MP84}.

In $G_2$ there are precisely 14 (conjugacy classes of rational) elements at which all the $C$-functions have integer values. In Table~\ref{efos} we list the 14 elements specified by the values of their coordinates $x\in F$, and show the values of $C$-functions at several lowest orbits of $W(G_2)$. Corresponding properties of the families $S$, $S^L$ and $S^S$ are not known.

\section{Preliminaries}\label{Prel}

The two simple roots of $G_2$ and their numbering are shown on the Dynkin diagram
\begin{center}
\parbox{.6\linewidth}
{\setlength{\unitlength}{1pt}
\def\kr{\circle{10}}
\def\cr{\circle*{10}}
\thicklines
\begin{picture}(20,30)
\put(88,24){$1$}
\put(90,14){\kr}
\put(86,0){$\alpha_1$}
\put(93,10){\line(1,0){18}}
\put(95,14){\line(1,0){12}}
\put(93,18){\line(1,0){18}}
\put(108,24){$2$}
\put(110,14){\cr}
\put(106,0){$\alpha_2$}
\end{picture}}
\end{center}
\medskip

\noindent
They span the real Euclidean space $\R^2$.
Standard conventions for writing the diagram identify the lengths and relative angle $\frac{5 \pi}{6}$ between the simple roots as given by
\begin{align*}
  \l\alpha_1,\alpha_1\r=2,\qquad
  \l\alpha_1,\alpha_2\r=-1,\qquad
  \l\alpha_2,\alpha_2\r=\frac{2}{3}\,.
\end{align*}

Relations between the four bases we will use can be written explicitly:
\begin{alignat*}{4}
\alpha_1&=2\omega_1-3\omega_2\,,&\qquad
\alpha_2&=-\omega_1+2\omega_2\,,&\qquad
\omega_1&=2\alpha_1+3\alpha_2\,,&\qquad
\omega_2&= \alpha_1+2\alpha_2\,,\\
\check\alpha_1&=2\check\omega_1-\check\omega_2\,,&\qquad
\check\alpha_2&=-3\check\omega_1+2\check\omega_2\,,&\qquad
\check\omega_1&=2\check\alpha_1+\check\alpha_2\,,&\qquad
\check\omega_2&=3\check\alpha_1+2\check\alpha_2\,.
\end{alignat*}
They follow from the $\Z$-duality requirement
$$
\l\alpha_j,\check\omega_k\r=\l\check\alpha_j,\omega_k\r=\delta_{jk}\,,\qquad j,k=1,2\,.
$$
The link between $\alpha$- and $\omega$-bases is provided by the Cartan matrix of $G_2$ and its inverse,
$$
 \begin{pmatrix}\alpha_1 & \alpha_2\end{pmatrix}
=\left(\begin{smallmatrix}2 & -3\\-1 & 2\end{smallmatrix}\right)
\begin{pmatrix}\omega_1 & \omega_2\end{pmatrix}\,,\qquad
\begin{pmatrix}\omega_1 & \omega_2\end{pmatrix}
=\left(\begin{smallmatrix}2 & 3\\1 & 2\end{smallmatrix}\right)
 \begin{pmatrix}\alpha_1 & \alpha_2\end{pmatrix}
$$
The determinant of the Cartan matrix is 1, therefore the root lattice $Q$ and the weight lattice $P$ coincide in the case of $G_2$.
$$
Q=P=\Z\alpha_1+\Z\alpha_2=\Z \omega_1+\Z \omega_2\,.
$$

Reflections $r_1$ and $r_2$ in mirrors passing through the origin and orthogonal to simple roots generate the root system $\Delta$ from the simple roots.

The set of positive roots $\Delta_+$, the set of positive long roots $\Delta^L_+$, the set of positive short roots $\Delta^S_+$ and their half sums $\rho$, $\rho^L$, $\rho^S$ are the following:
\begin{alignat*}{2}
\Delta_+ &=\{\alpha_1,\alpha_2,\alpha_1+\alpha_2,\alpha_1+2\alpha_2,
        \alpha_1+3\alpha_2,2\alpha_1+3\alpha_2\},&\quad
\rho&=3\alpha_1+5\alpha_2=\omega_1+\omega_2\,,\\
\Delta^L_+ &=\{\alpha_1,\ \alpha_1+3\alpha_2,\ 2\alpha_1+3\alpha_2\},&\quad
\rho^L&=2\alpha_1+3\alpha_2=\omega_1\,,\\
\Delta^S_+ &=\{\alpha_2,\ \alpha_1+\alpha_2,\ \alpha_1+2\alpha_2\},&\quad
\rho^S&=\alpha_1+2\alpha_2=\omega_2\,.
\end{alignat*}

\begin{figure}
\centering
\includegraphics{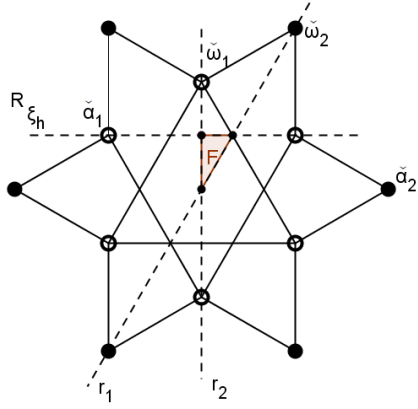}
 \caption{{\footnotesize A schematic view of the co-root system of $G_2$. A shaded triangle is a fundamental region $F$. Black (open) dots are the long (short) co-roots.}}
\label{g2}
\end{figure}

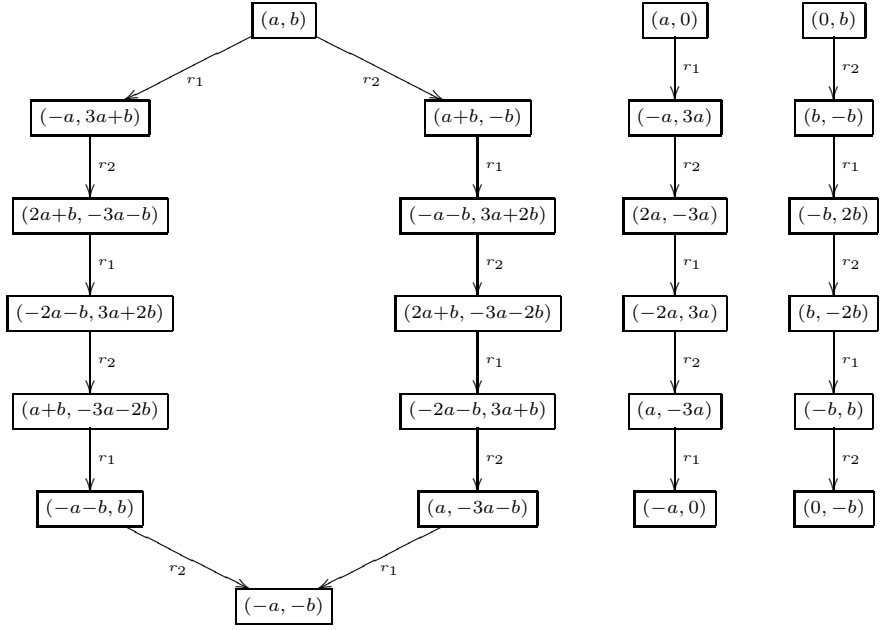
\begin{figure}[h]
{\scriptsize $\xymatrix{& *+[F]{(a,b)}\ar[dl]^{r_1} \ar[dr]_{r_2}&& *+[F]{(a,0)}\ar[d]^{r_1} &   *+[F]{(0,b)}\ar[d]^{r_2}\\
 *+[F]{({-}a,3a{+}b)} \ar[d]^{r_2}& & *+[F]{(a{+}b,{-}b)} \ar[d]^{r_1} &  *+[F]{({-}a,3a)} \ar[d]^{r_2}&     *+[F]{(b,{-}b)} \ar[d]^{r_1}\\
*+[F]{(2a{+}b,{-}3a{-}b)}\ar[d]^{r_1} & & *+[F]{({-}a{-}b,3a{+}2b)}\ar[d]^{r_2}& *+[F]{(2a,{-}3a)}\ar[d]^{r_1} & *+[F]{({-}b,2b)}\ar[d]^{r_2}\\
*+[F]{({-}2a{-}b,3a{+}2b)}\ar[d]^{r_2} & & *+[F]{(2a{+}b,{-}3a{-}2b)}\ar[d]^{r_1} & *+[F]{({-}2a,3a)}\ar[d]^{r_2}   &  *+[F]{(b,{-}2b)}\ar[d]^{r_1}\\
 *+[F]{(a{+}b,{-}3a{-}2b)}\ar[d]^{r_1} & & *+[F]{({-}2a{-}b,3a{+}b)}\ar[d]^{r_2}&  *+[F]{(a,{-}3a)}\ar[d]^{r_1}   & *+[F]{({-}b,b)}\ar[d]^{r_2}\\
*+[F]{({-}a{-}b,b)}\ar[dr]_{r_2} & & *+[F]{(a,{-}3a{-}b)}\ar[dl]^{r_1} &  *+[F]{({-}a,0)}    & *+[F]{(0,{-}b)}\\
& *+[F]{({-}a,{-}b)} &  &}$}
\medskip
\caption{{\footnotesize Weyl group orbits of the three types of dominant points $\lambda=(a,b)$, $(a,0)$ and $(0,b)$.}}\label{orbits}
\end{figure}

\section{Orbit functions of $G_2$}\label{OF}
\subsection{Weyl group orbits of $G_2$}\

The Weyl group $W(G_2)$, which is generated by two reflections $r_{\alpha_1}\equiv r_1$ and $r_{\alpha_2}\equiv r_2$, is abstractly defined by the following conditions on its generating reflections:
\begin{gather}\label{presentation}
r_1^2=r_2^2=1\,,\qquad
\left(r_1r_2\right)^6=1\,.
\end{gather}

In addition, it is assumed that the exponent 6 in \eqref{presentation} is the lowest one that would allow the equality to hold.

Reflections \eqref{presentation} act in the 2-dimensional real Euclidean space $\R^2$ spanned by the simple roots, according to
\begin{gather}\label{reflection}
r_kx=x-\frac{2\l\alpha_k,x\r}{\l\alpha_k,\alpha_k\r}\alpha_k\,,\qquad
   x\in\R^2,\quad  k=1,2.
\end{gather}
Together with the affine reflection in the highest root $\xi=2\alpha_1+3\alpha_2
=\omega_1=\check\omega_1$,
\begin{gather}\label{affreflection}
R_\xi x=\xi+r_\xi x=\xi+x-\frac{2\l\xi,x\r}{\l\xi,\xi\r}\xi\,,
\end{gather}
\eqref{reflection} and \eqref{affreflection} generate the affine Weyl group of $G_2$.

A Weyl group orbit $W_\lambda$ is the set of distinct points (`weights') obtained from single seed point $\lambda$ by repeated reflections of \eqref{reflection}.
The points comprising an orbit are shown on Figure~\ref{orbits}. Note three properties of the orbits of $W(G_2)$ that are of importance to us: \newline
(i) Any point $x\in F$ belongs to only one  orbit $W_\lambda$.\newline
(ii) There is one point, called dominant, $\lambda\in W_\lambda$ with non-negative coordinates in $\omega$-basis. \newline
(iii) For each $\mu\in W_\lambda$ there is also the point $-\mu\in W_\lambda$.

The fundamental domain $F$ of $G_2$ is the convex hull of its three vertices
\begin{gather}\label{domainF}
F=\{0,\tfrac12\check\omega_1,\tfrac13\check\omega_2\}.
\end{gather}
There is a one-to-one correspondence between dominant points of the orbits and the points of $F$.

\subsection{The `new' and `old' orbit functions}\

In general, the function of an orbit $W_\lambda$ of $W(G_2)$ is understood here to be the sum of exponential terms
\begin{equation}\label{OF}
\sum_{\mu\in W_\lambda}\sigma(\mu)e^{2\pi i\l\mu,x\r}\,,\qquad
    \lambda\in P^+,\quad x\in\R^2,\quad\sigma(\mu)=\pm1
\end{equation}
where the summation extends over the points $\mu$ of the Weyl group orbit $W_\lambda$ containing the dominant point $\lambda$. The homomorphism
\begin{gather*}\label{homo}
\sigma: W \longrightarrow \{\pm 1\}
\end{gather*}
determines the four families of orbit functions \cite{MMP}.

The map $\sigma(w)$ is given by the values $\sigma(r_i) \in \{\pm 1\}$, $i=1,2$. For $G_2$, the necessary and sufficient condition for a homomorphism to exist is that $(\sigma(r_1)\sigma(r_2))^6 =1$, which is automatically satisfied. Consequently there are $4$ homomorphisms $\sigma$ and their orbit functions:
\begin{equation*}\label{signs}
\begin{array}{ll}
\sigma(r_1)=\sigma(r_2)=1&\qquad\Longrightarrow\quad C\\
\sigma(r_1)=\sigma(r_2)=-1&\qquad\Longrightarrow\quad S\\
\sigma(r_1)=-1,\quad \sigma(r_2)=1&\qquad\Longrightarrow\quad S^L\\
\sigma(r_1)=1,\quad \sigma(r_2)=-1&\qquad\Longrightarrow\quad S^S
\end{array}
\end{equation*}
The homomorphisms leading to $C$- and $S$-functions are `old' \cite{P,PZ2,PZ3}, while the remaining two leading to the $S^L$- and $S^S$-functions are `new' \cite{MMP}.

Signs of individual exponential terms in orbit functions are given in Table~\ref{signs}.

\subsection{Characters of $G_2$ and the orbit functions}\

The character $\chi_\lambda(x)$ is given either by summation over the weight system of the irreducible representation with the highest weight $\lambda$ or by the Weyl character formula,
\begin{gather}\label{character}
\chi_\lambda(x)=\sum_\mu m_{\lambda\mu}C_\mu(x)=
\frac{S_{\lambda+\rho}(x)}{S_{\rho}(x)}\,, \qquad \lambda,\mu\in P^+, \quad \rho=(1,1), \quad x\in \R^2
\end{gather}
where the summation extends over the dominant weights $\mu$ in the weight system of the representation $\lambda$.

The $C$-function $C_{\mu}(x)$ is defined by summation over the points $\eta$ of one $W$-orbit, appearing in the scalar products $\l\eta,x\r$,
\begin{gather}\label{Cfunc}
C_{\mu}(x)=\sum_{\eta\in W_\mu}e^{2\pi i\l\eta,x\r}\,.
\end{gather}
The coefficient $m_{\lambda\mu}$ in \eqref{character} is the multiplicity of the dominant weight $\mu$ in the weight system of the irreducible representation specified by its highest weight $\lambda$.
The multiplicities are calculated using algorithm \cite{MP82}, or they can be read from the extensive tables \cite{BMP}. By a suitable ordering of the weights in $P^+$, used for example in \cite{BMP}, the matrix $(m_{\lambda\mu})$ becomes triangular with all diagonal entries being 1. Such a matrix can be inverted into $(m_{\lambda\mu}^{-1})$. Therefore, in addition to \eqref{character}, we always have also
\begin{gather}\label{invertedcharacter}
C_\mu(x)=\sum_\lambda m_{\lambda\mu}^{-1}\chi_\lambda(x)
   =\frac1{S_\rho (x)}\sum_\lambda m_{\lambda\mu}^{-1} S_{\lambda+\rho}(x)\,.
\end{gather}
Many known properties of characters carry over through \eqref{invertedcharacter} into properties of the functions of the $C$ family.
\begin{example}\

Reading $(m_{\lambda\mu})$ as the $8\times8$ matrix from \cite{BMP}, and inverting that matrix, we get in particular $C_{(2,1)}(x)$ as the following linear combination of the seven lowest characters
$$
C_{(2,1)}(x)=\chi_{(2,1)}(x)-\chi_{(0,2)}(x)-\chi_{(3,0)}(x)
            -\chi_{(1,1)}(x)+\chi_{(2,0)}(x)+\chi_{(0,1)}(x).
$$
\end{example}
\medskip
In principle, all irreducible characters of $G_2$ are also obtained from the character generating function \cite{GS} of $G_2$. Expanding the generating function into power series, the characters are found as coefficients of appropriate powers of the series.

An interesting comparison of the new functions with the character  $\chi_\lambda(x)$ is made when they are written in the form closely resembling \eqref{character},
\begin{gather}\label{ratios}
\begin{aligned}
\chi^L_\lambda(x)=\frac{S^L_{\lambda+\rho^L}(x)}{S^L_{\rho^L}(x)}
  =\sum_{\mu\in W_{\lambda}}m^L_{\lambda,\mu}C_\mu(x) \,, \\
\chi^S_\lambda(x)=\frac{S^S_{\lambda+\rho^S}(x)}{S^S_{\rho^S}(x)}
  =\sum_{\mu\in W_{\lambda}}m^S_{\lambda,\mu}C_\mu(x)\,.
\end{aligned}
\end{gather}
The symbols  $\rho^L=(1,0)$ and $\rho^S=(0,1)$ stand for the half sums of respectively positive long roots and positive short roots in $\omega$-basis.

While the character $\chi_\lambda(x)$ is probably the most important function in representation theory, no role of $\chi^L_\lambda(x)$ and $\chi^S_\lambda(x)$ has been identified so far. Unlike the coefficients $m_{\lambda\mu}$ in \eqref{character}, no analogous algorithm to \cite{MP82} is known for coefficients $m^L_{\lambda,\mu}$ and $m^S_{\lambda,\mu}$, but their values can be readily found for the lowest few dominant weights $\lambda$ using the recursion relations for the appropriate orbit functions, see section~\ref{Decomp}.

\begin{example}\

In order to illustrate how different the functions
$\chi_\lambda(x)$, $\chi^L_\lambda(x)$ and $\chi^S_\lambda(x)$ really are for the same value of $x$, let us compare the lowest few characters \eqref{character} with the corresponding expressions for \eqref{ratios}.
\begin{align*}
\chi_{(1,0)}   &=C_{(1,0)}+ C_{(0,1)}+2C_{(0,0)},\\
\chi^L_{(1,0)} &=C_{(1,0)}+2C_{(0,0)},\\
\chi^S_{(1,0)} &=C_{(1,0)}+2C_{(0,1)}+2C_{(0,0)},\\
\chi_{(0,1)} &=C_{(0,1)}+C_{(0,0)},\\
\chi^L_{(0,1)} &=C_{(0,1)},\\
\chi^S_{(0,1)} &=C_{(0,1)}+2C_{(0,0)},\\
\chi_{(1,1)}   &=C_{(1,1)}+2C_{(0,2)}+2C_{(1,0)}+4C_{(0,1)}+4C_{(0,0)},\\
\chi^L_{(1,1)} &=C_{(1,1)}+C_{(0,2)}+2C_{(0,1)},\\
\chi^S_{(1,1)} &=C_{(1,1)}+2C_{(0,2)}+3C_{(1,0)}+4C_{(0,1)}+4C_{(0,0)},\\
\chi_{(2,0)}   &=C_{(2,0)}+C_{(0,3)}+C_{(1,1)}+2C_{(0,2)}+3C_{(1,0)}+3C_{(0,1)}+5C_{(0,0)},\\
\chi^L_{(2,0)} &=C_{(2,0)}+C_{(0,3)}+2C_{(1,0)}+3C_{(0,0)},\\
\chi^S_{(2,0)} &=C_{(2,0)}+C_{(1,1)}+2C_{(0,2)}+2C_{(1,0)}+2C_{(0,1)}+2C_{(0,0)},\\
\chi_{(0,2)}   &=C_{(0,2)}+C_{(1,0)}+2C_{(0,1)}+3C_{(0,0)},\\
\chi^L_{(0,2)} &=C_{(0,2)}+C_{(0,1)},\\
\chi^S_{(0,2)} &=C_{(0,2)}+C_{(1,0)}+2C_{(0,1)}+3C_{(0,0)},\\
\chi_{(0,3)}   &=C_{(0,3)}+C_{(1,1)}+2C_{(0,2)}+3C_{(1,0)}+4C_{(0,1)}+5C_{(0,0)},\\
\chi^L_{(0,3)} &=C_{(0,3)}+2C_{(1,0)}+2C_{(0,0)},\\
\chi^S_{(0,3)} &=C_{(0,3)}+C_{(0,1)}+C_{(1,1)}+2C_{(0,2)}+2C_{(1,0)}+2C_{(0,1)}+4C_{(0,0)}.\\
\end{align*}
\end{example}

The particular cases of characters with fixed value of $x\in F$ and general dominant weight $\lambda$ are also useful. For example, the dimension of an irreducible representation of $G_2$ with the highest weight $\lambda=(a,b)\in P^+$ is given by the following:
\begin{gather}\label{dimension}
\chi_{(a,b)}(0,0)= \tfrac1{120}(a+1)(b+1)(a+b+2)(2a+b+3)(3a+b+4)(3a+2b+5)\,,
\notag\\
a,b\in\Z^{\geq0}.
\end{gather}
Analogous expressions for $\chi^L_{(a,b)}(0,0)$ and $\chi^S_{(a,b)}(0,0)$ are yet to be found and, more importantly, their interpretation should be given.

Combining the orbits, Figure~\ref{orbits}, and the signs \eqref{signs}, one finds also
\begin{gather*}
\begin{aligned}
S^L_{(a,b)}(0,0)=0\,, \qquad
S^L_{(a,0)}(0,0)=0\,, \qquad
S^L_{(0,b)}(0,0)=0\,, \\
S^S_{(a,b)}(0,0)=0\,, \qquad
S^S_{(a,0)}(0,0)=0\,, \qquad
S^S_{(0,b)}(0,0)=0\,,
\end{aligned}\qquad a,b\in\Z^{>0}.
\end{gather*}

\medskip

\subsection{Explicit form of the orbit functions}\

It is convenient to specify the general functions of types $S$, $S^L$ and $S^S$ by their dominant point written as the sum of two terms, one general and one being the half sum of the corresponding  positive roots. More precisely,
\begin{gather*}
\begin{aligned}
S_{\lambda+\rho}(x)&=S_{\lambda+(1,1)}(x)
 =S_{(1,1)}(x)\sum_{\mu\in W_{\lambda}}m_{\lambda,\mu}
 C_\mu(x),\\
S^L_{\lambda+\rho^L}(x)&=S^L_{\lambda+(1,0)}(x)
 =S^L_{(1,0)}(x)\sum_{\mu\in W_{\lambda}}m^L_{\lambda,\mu}
 C_\mu(x),\\
S_{\lambda+\rho^S}(x)&=S^S_{\lambda+(0,1)}(x)
 =S^S_{(0,1)}(x)\sum_{\mu\in W_{\lambda}}m^S_{\lambda,\mu}
 C_\mu(x)
\end{aligned}
\end{gather*}
where $\lambda,\mu\in P^+$ and $x\in\R^2$. Substitution of the orbit points of Figure ~\ref{orbits} into \eqref{Cfunc}, yields
\begin{align*}
C_{(a,b)}(x)&=2[\cos2\pi \l(a,b),x\r + \cos2\pi \l(a+b,-b),x\r\notag\\
    &+ \cos2\pi \l(-a,3a+b),x\r +   \cos2\pi \l(-a-b,3a+2b),x\r\\
    &+ \cos2\pi \l(2a+b,-3a-2b),x\r+\cos2\pi \l(-2a-b,3a+b),x\r],\\\notag
C_{(a,0)}(x)&= 2[\cos(2\pi\l(a,0),x\r)+\cos(2\pi\l({-}a,3a),x\r)
      +\cos(2\pi\l(2a,{-}3a),x\r)], \\
C_{(0,b)}(x)&=2[\cos(2\pi\l(0,b),x\r)+\cos(2\pi\l(b,{-}b),x\r)+\cos(2\pi\l({-}b,2b),x\r)].
\end{align*}
Choosing a suitable basis for $x\in\R^2$, it remains to calculate the scalar products in each cosine.

Another way to explicitly describe the orbit functions is to supply the factors $\pm1$ that multiply exponential terms in the orbit function \eqref{OF} for $C$-, $S$-, $S^L$- and $S^S$-functions. In the case of $C$-functions, all signs are positive. Table~\ref{signs} presents a concise way to provide the signs in other cases.

\begin{table}[h]
{\footnotesize
  \begin{tabular}{|c||c|c|c|c|}
    \hline \rule{0pt}{10pt}
    $\lambda=(a,b)_{\o}$ & $C$ & $S$ & $S^S$ & $S^L$\rule{0pt}{10pt}\\[2pt]
    \hline\hline \rule{0pt}{10pt}
    $(a,b)$ & $+$ & $+$ & $+$ & $+$\rule{0pt}{10pt}\\[2pt]
    \hline \rule{0pt}{10pt}
    $({-}a,3a{+}b)$ & $+$ & $-$ & $+$ & $-$\rule{0pt}{10pt}\\[2pt]
    \hline \rule{0pt}{10pt}
    $(a{+}b,{-}b)$ & $+$ & $-$ & $-$ & $+$\rule{0pt}{10pt}\\[2pt]
    \hline \rule{0pt}{10pt}
    $(2a{+}b,{-}3a{-}b)$ & $+$ & $+$ & $-$ & $-$\rule{0pt}{10pt}\\[2pt]
    \hline \rule{0pt}{10pt}
    $({-}a{-}b,3a{+}2b)$ & $+$ & $+$ & $-$ & $-$\rule{0pt}{10pt}\\[2pt]
    \hline \rule{0pt}{10pt}
    $({-}2a{-}b,3a{+}2b)$ & $+$ & $-$ & $-$ & $+$\rule{0pt}{10pt}\\[2pt]
    \hline \rule{0pt}{10pt}
    $(2a{+}b,{-}3a{-}2b)$ & $+$ & $-$ & $+$ & $-$\rule{0pt}{10pt}\\[2pt]
    \hline \rule{0pt}{10pt}
    $(a{+}b,{-}3a{-}2b)$ & $+$ & $+$ & $+$ & $+$\rule{0pt}{10pt}\\[2pt]
    \hline \rule{0pt}{10pt}
    $({-}2a{-}b,3a{+}b)$ & $+$ & $+$ & $+$ & $+$\rule{0pt}{10pt}\\[2pt]
    \hline \rule{0pt}{10pt}
    $({-}a{-}b,b)$ & $+$ & $-$ & $+$ & $-$\rule{0pt}{10pt}\\[2pt]
    \hline \rule{0pt}{10pt}
    $(a,{-}3a{-}b)$ & $+$ & $-$ & $-$ & $+$\rule{0pt}{10pt}\\[2pt]
    \hline \rule{0pt}{10pt}
     $({-}a,{-}b)$ & $+$ & $+$ & $-$ & $-$\rule{0pt}{10pt}\\[2pt]
    \hline
  \end{tabular} \qquad\begin{tabular}{c}
    \begin{tabular}{|c||c|c|}
    \hline \rule{0pt}{10pt}
    $\lambda=(a,0)_{\o}$ & $C$ & $S^L$\rule{0pt}{10pt}\\[2pt]
    \hline\hline \rule{0pt}{10pt}
    $(a,0)$ & $+$& $+$\rule{0pt}{10pt}\\[2pt]
    \hline \rule{0pt}{10pt}
    $({-}a,3a)$ & $+$ & $-$\rule{0pt}{10pt}\\[2pt]
    \hline \rule{0pt}{10pt}
    $(2a,{-}3a)$ & $+$ & $-$\rule{0pt}{10pt}\\[2pt]
    \hline \rule{0pt}{10pt}
    $({-}2a,3a)$ & $+$ & $+$\rule{0pt}{10pt}\\[2pt]
    \hline \rule{0pt}{10pt}
    $(a,{-}3a)$ & $+$ & $+$\rule{0pt}{10pt}\\[2pt]
    \hline \rule{0pt}{10pt}
    $({-}a,0)$ & $+$ & $-$\rule{0pt}{10pt}\\[2pt]
    \hline
  \end{tabular} \\
  \ \\
  \
  \\
 \begin{tabular}{|c||c|c|}
    \hline \rule{0pt}{10pt}
    $\lambda=(0,b)_{\o}$ & $C$ & $S^S$ \rule{0pt}{10pt}\\[2pt]
    \hline\hline \rule{0pt}{10pt}
    $(0,b)$ & $+$ & $+$\rule{0pt}{10pt}\\[2pt]
    \hline \rule{0pt}{10pt}
    $(b,{-}b)$ & $+$ & $-$ \rule{0pt}{10pt}\\[2pt]
    \hline \rule{0pt}{10pt}
    $({-}b,2b)$ & $+$ & $-$ \rule{0pt}{10pt}\\[2pt]
    \hline \rule{0pt}{10pt}
    $(b,{-}2b)$ & $+$ & $+$ \rule{0pt}{10pt}\\[2pt]
   \hline \rule{0pt}{10pt}
    $({-}b,b)$ & $+$ & $+$\rule{0pt}{10pt}\\[2pt]
    \hline \rule{0pt}{10pt}
    $(0,{-}b)$ & $+$ & $-$ \rule{0pt}{10pt}\\[2pt]
    \hline
  \end{tabular}\end{tabular}}
\medskip
  \caption{{\footnotesize Description of the orbit functions of types $C$, $S$, $S^L$ and $S^S$ of an orbit $W_\lambda$. Each row pertains to the exponential term in the orbit function \eqref{OF}, that carries the point $\mu$ shown in the first column.  The subsequent four columns provide the signs $\sigma(\mu)$ of the same term in each of the four orbit functions.}}\label{signs}
\end{table}

\subsection{Orbit functions at the boundaries of $F$}\

It is known and easy to verify directly for $G_2$ that all $C$-functions are symmetric with respect to reflections in the boundaries of $F$, while all $S$-functions are antisymmetric with respect to such reflections.

Geometrically, domain $F$ of $G_2$, see \eqref{domainF}, is half of an equilateral triangle, its right angle being at the vertex $\tfrac12\check\omega_1$. The sides of $F$, which are of different length, can be distinguished by the reflection mirror they are part of. Specifically,
\begin{alignat*}{2}
r_1\ &:\   \text{ reflection
         across the side}&\quad &0,\tfrac13\check\omega_2\notag\\
r_2\ &:\   \text{ reflection
         across the side}&\quad &0,\tfrac12\check\omega_1\\
r_\xi\ &:\ \text{ (affine) reflection
         across the side}&\quad &\tfrac12\check\omega_1,\tfrac13\check\omega_2 \notag
\end{alignat*}

The orbit function, being continuous and with all continuous derivatives, vanish at the boundaries of $F$ in the following cases:
\begin{align}\label{atboundaries}
\text{orbit functions $S$ and $S^L$ are antisymmetric at}\quad
           &0,\tfrac13\check\omega_2\notag\\
\text{orbit functions $S$ and $S^S$ are antisymmetric at} \quad
           &0,\tfrac12\check\omega_1\\
\text{orbit functions $S$ and $S^L$ are antisymmetric at} \quad
          &\tfrac12\check\omega_1,\tfrac13\check\omega_2\notag
\end{align}
At other boundaries of $F$, the orbit functions are symmetric, i.e. their normal derivative to the boundary is zero.

\subsection{Continuous orthogonality of the orbit functions of $G_2$}\

Orthogonality of the irreducible characters over $F$ is well known  \cite{BS}. Also the orthogonality of $C$- and $S$-functions has been shown \cite{MP06} in general. An demonstration of orthogonality for all orbit functions, including the new ones, follows a similar path.

Since a $C$-function can be viewed as a linear combination of irreducible characters, the orthogonality of the $C$-function follows. A product of a pair $S^LS^L$ of functions is a product of linear combinations of $C$-functions multiplied by the particular $S^L_{\rho^L}S^L_{\rho^L}$, which decomposes into the $C$-functions (see Table~\ref{products}). Since products of $C$-functions are completely decomposable, the orthogonality of $S^L$-functions follows. An analogous argument can be made for the $S^S$-functions. It remains only to work out what the value of products is when the same functions are multiplied.

In the case of $G_2$, the orthogonality of $C$-functions was described in \cite{PZ2}, and the orthogonality of $S$-functions is found in \cite{PZ3}. We recall and complete those results with the orthogonality relations of $S^L$- and $S^S$-functions. In fact, corresponding double integrals can be calculated directly.

A continuous scalar product for any two square integrable functions $f(x)$ and $g(x)$ is defined on the fundamental region $F$ of $G_2$:
\begin{equation}\l f(x),g(x)\r  :=\int\limits_F f(x)\overline{g(x)}\textrm{d}F, \qquad x=(x_1,x_2).\end{equation}
For $\lambda=(a,b)$, $\mu=(c,d)$,  $(a,b,c,d\geq0)$, we have the $G_2$ continuous orthogonality relations:

\begin{gather*}
  \l C_\lambda,C_\mu\r = \sqrt{3} \int\limits_0^\frac{1}{2}\int\limits_0^{\frac{1}{3}-\frac{2}{3}x_1} C_\lambda(x)\overline{C_\mu(x)}\textrm{d}x_2\textrm{d}x_1 =\\
 \qquad =  \left\{\begin{array}{ll}
\frac{\sqrt 3}{12}, & \textrm{if } a=b=c=d=0,\\
\frac{\sqrt 3}{2}, & \textrm{if } a=c>0 \ \mathrm{and} \ b=d=0\\
 & \quad \mathrm{or} \ a=c=0 \ \mathrm{and} \ b=d>0, \\
 \sqrt 3, & \textrm{if } a=c>0 \ \mathrm{and} \ b=d>0,\\
0, & \textrm{in other cases.}
  \end{array}\right.
\end{gather*}

\begin{gather*}
  \l S_{(a,b)},S_{(c,d)}\r =
 \left\{\begin{array}{ll}
 \sqrt 3, & \textrm{if } a=c>0 \ \mathrm{and} \ b=d>0,\\
 0, & \textrm{in other cases}.
  \end{array}\right.
\end{gather*}

\begin{gather*}
  \l S^L_{(a,b)},S^L_{(c,d)}\r =
 \left\{\begin{array}{ll}
\sqrt 3, & \textrm{if } a=c>0 \ \mathrm{nd} \ b=d>0,\\
 \frac{\sqrt 3}{2}, & \textrm{if } a=c>0 \ \mathrm{and} \ b=d=0,\\
 0, & \textrm{in other cases}.
  \end{array}\right.
\end{gather*}

\begin{gather*}
  \l  S^S_{(a,b)},S^S_{(c,d)}\r =
  \left\{\begin{array}{ll}
\sqrt 3, & \textrm{if } a=c>0 \ \mathrm{and} \ b=d>0, \\
 \frac{\sqrt 3}{2}, & \textrm{if } a=c=0 \ \mathrm{and} \ b=d>0,\\
 0, & \textrm{in other cases}.
  \end{array}\right.
\end{gather*}

\section{Discretization of orbit functions of $G_2$.}\label{Discret}

Discretization of the $C$-functions of any compact simple Lie group was invented \cite{MP87} as a practically indispensable tool for solving large problems in representation theory, see for example \cite{GP}. We present it here because the discretization of orthogonal polynomials in \cite{PSz} would not be possible without discretized orbit functions.

The main steps of discretization are\newline
(i) Setting up lattice fragment $F_M\subset F$ of desired density determined by our choice of a natural number $M$, see Figure~\ref{grids}. That is, describing points $x\in F_M$. \newline
(ii) Sampling the orbit functions at the points of $F_M$. \newline
(iii) Demonstrating the orthogonality of the orbit functions when summed up over $F_M$ with the weighting function $c_x$ appropriate for each point $x\in F_M$.

\subsection{The lattice grid $F_M$.}\

The lattice grid $F_M=\frac1M \check P\cap F$ is a finite fragment of the weight lattice $\check P$ of $G_2$, refined by our choice of positive integer $M$. The same grid is used for all four types of the orbit functions.

Suppose that a positive integer $M$ has been chosen and fixed. The points of $F_M$ are in one-to-one correspondence with the triples $[s_0,s_1,s_2]$ such that
\begin{gather}\label{sumrule}
M=s_0+2s_1+3s_2\,,\qquad s_0,s_1,s_2\in\Z^{\geq0}\,.
\end{gather}
Each triple is then drawn as the point
\begin{gather*}
\tfrac{s_1}{M}\check\omega_1+\tfrac{s_2}{M}\check\omega_2
=\left(\tfrac{s_1}{M},\tfrac{s_2}{M}\right)\in F_M\,.
\end{gather*}
Thus points of $F_M$ are determined by the sum rule \eqref{sumrule}. In particular, the triple $[M,0,0]$ is the vertex of $F$ that is the origin of $\check P$ for all choices of $M$.

\begin{example}\

We write some examples of the points of $F_M$ together with their positions in $F$ shown on Figure ~\ref{grids}. A point is given in terms of its Kac coordinates $[s_0,s_1,s_2]$ as well as by its coordinates in the $\check\omega$-basis:
\begin{align*}
&M=2\ :&\quad
&[2,0,0]=(0,0),&\quad
&[0,1,0]=(\tfrac12,0).
&&
\\
&M=3\ :&\quad
&[3,0,0]=(0,0),&\quad
&[1,1,0]=(\tfrac13,0),&\quad
&[0,0,1]=(0,\tfrac13).
\\
&M=6\ :&\quad
&[6,0,0] =(0,0)\,,&\quad
&[4,1,0] =(\tfrac16,0)\,,&\quad
&[3,0,1] =(0,\tfrac16)\,,
\\
&&&[2,2,0] =(\tfrac13,0)\,,&\quad
&[1,1,1] =(\tfrac16,\tfrac16)\,,&\quad
&[0,3,0] =(\tfrac12,0)\,,
\\
&&&[0,0,2] =(0,\tfrac13)\,.&&&&
\\
&M=10\ :                    & \quad
&[10,0,0]=(0,0)\,,          & \quad
&[8,1,0] =(\tfrac1{10},0)\,,& \quad
&[6,2,0] =(\tfrac2{10},0)\,,
\\
&&&[4,3,0] =(\tfrac3{10},0)\,,& \quad
  &[2,4,0] =(\tfrac4{10},0)\,,& \quad
  &[0,5,0] =(\tfrac5{10},0)\,,& \quad
\\
&&&[7,0,1] =(0,\tfrac1{10})\,,         & \quad
  &[4,0,2] =(0,\tfrac2{10})\,,          & \quad
  &[1,0,3] =(0,\tfrac3{10})\,,          & \quad
\\
&&&[5,1,1] =(\tfrac1{10},\tfrac1{10})\,,& \quad
  &[3,2,1]=(\tfrac2{10}, \tfrac1{10})\,,& \quad
  &[1,3,1] =(\tfrac3{10},\tfrac1{10})\,,& \quad
\\
&&&[2,1,2] =(\tfrac1{10},\tfrac2{10})\,,& \quad
  &[0,2,2] =(\tfrac2{10},\tfrac2{10})\,.
\end{align*}

\begin{figure}[h]
\centering
\includegraphics[scale=0.3]{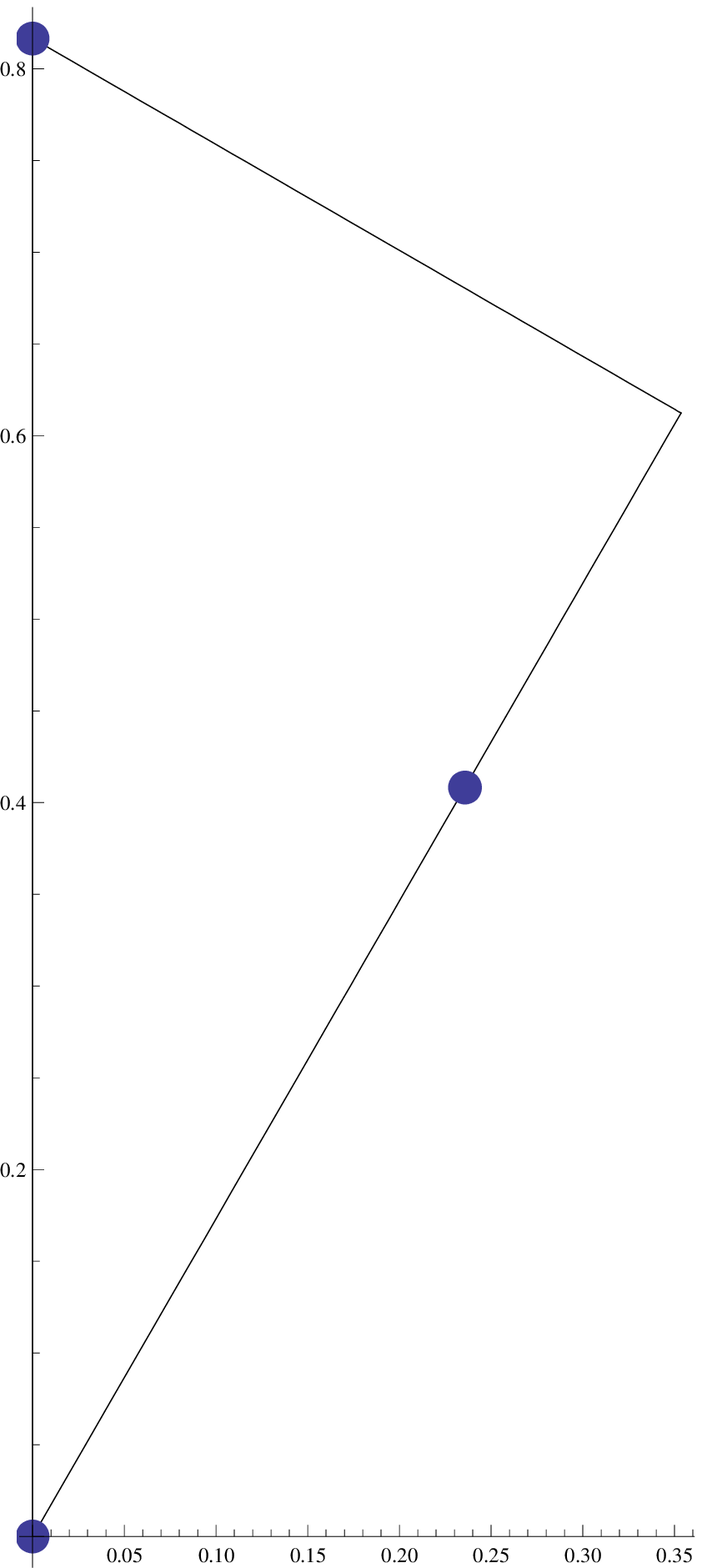} \qquad \qquad \includegraphics[scale=0.3]{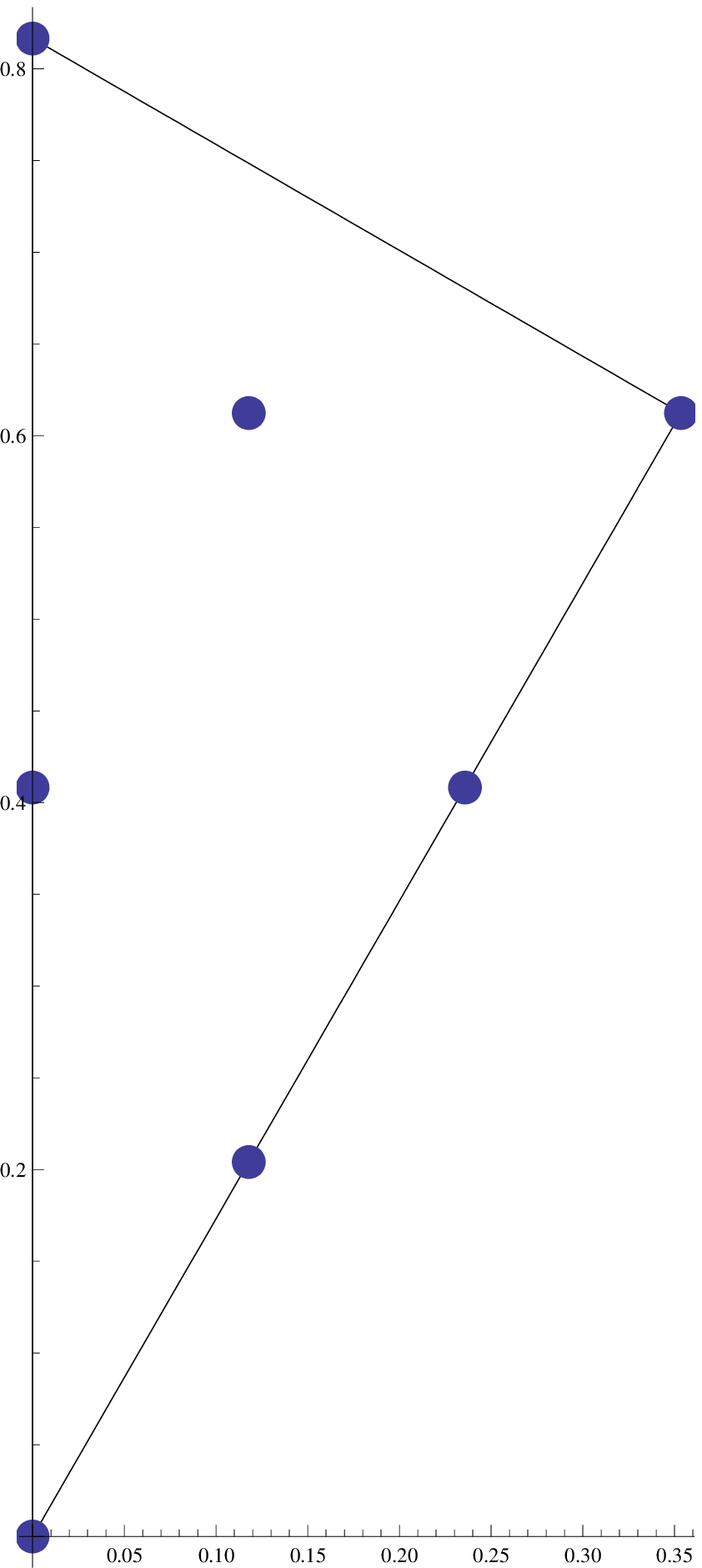}  \qquad \qquad
\includegraphics[scale=0.3]{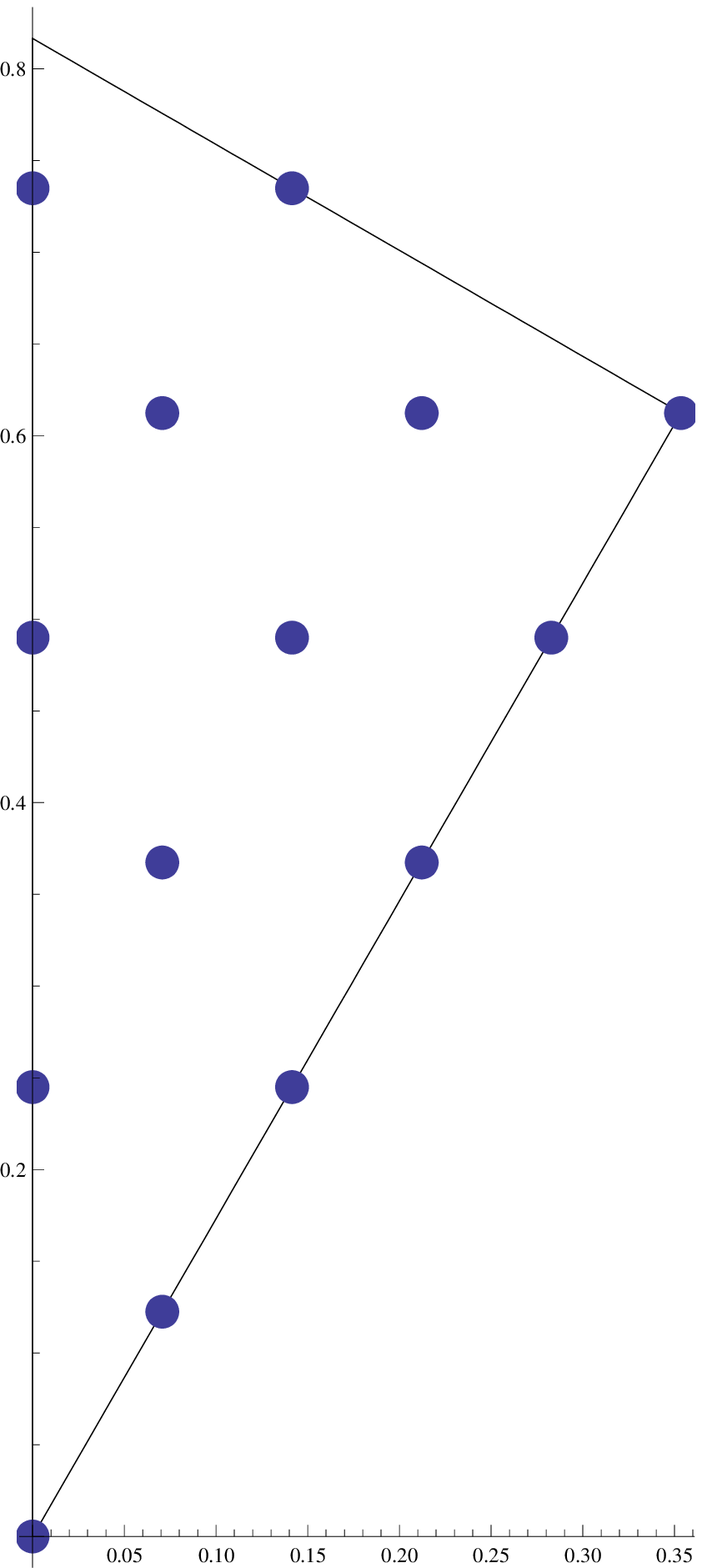}\\
$M=3$ \hskip 3 cm $M=6$ \hskip 3 cm $M=10$\\
\includegraphics[scale=0.3]{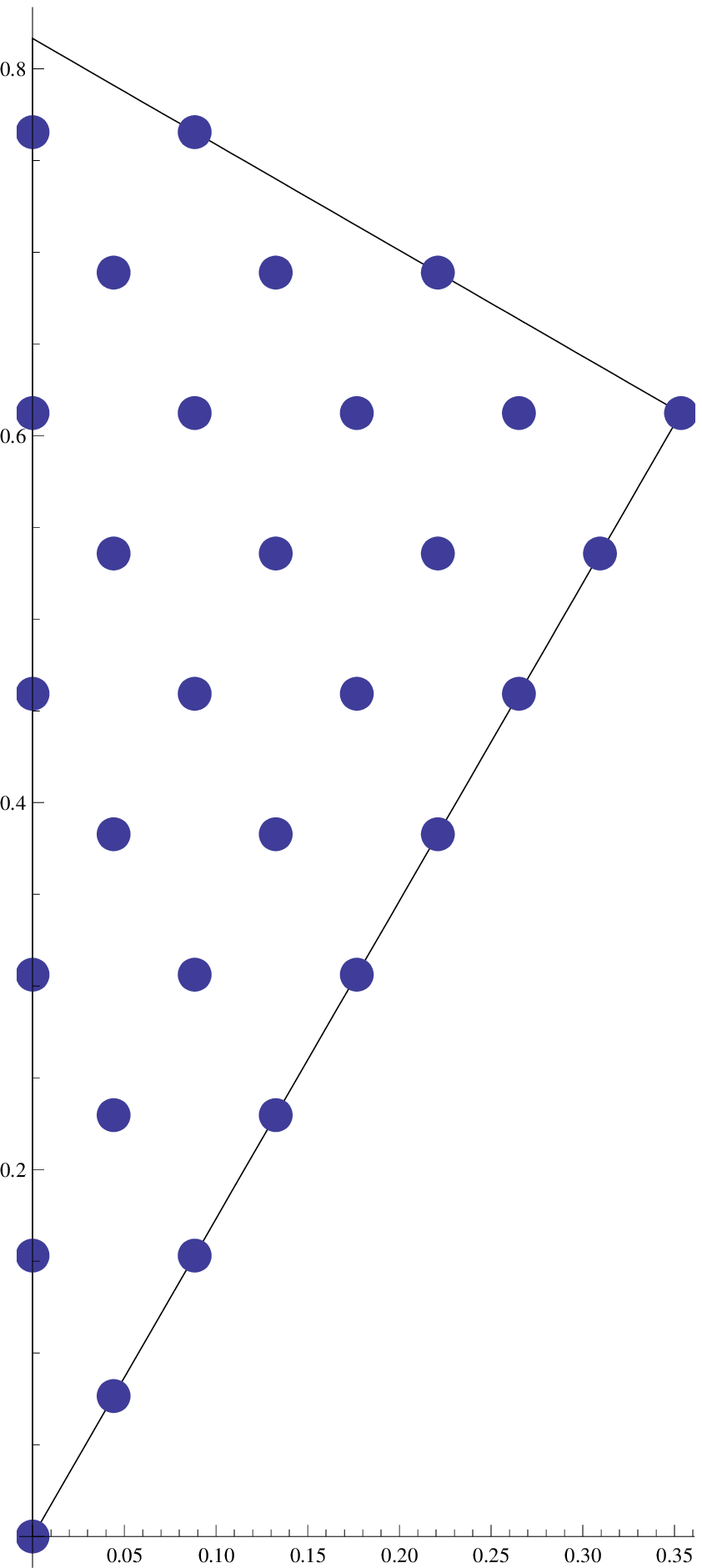} \qquad \qquad \includegraphics[scale=0.3]{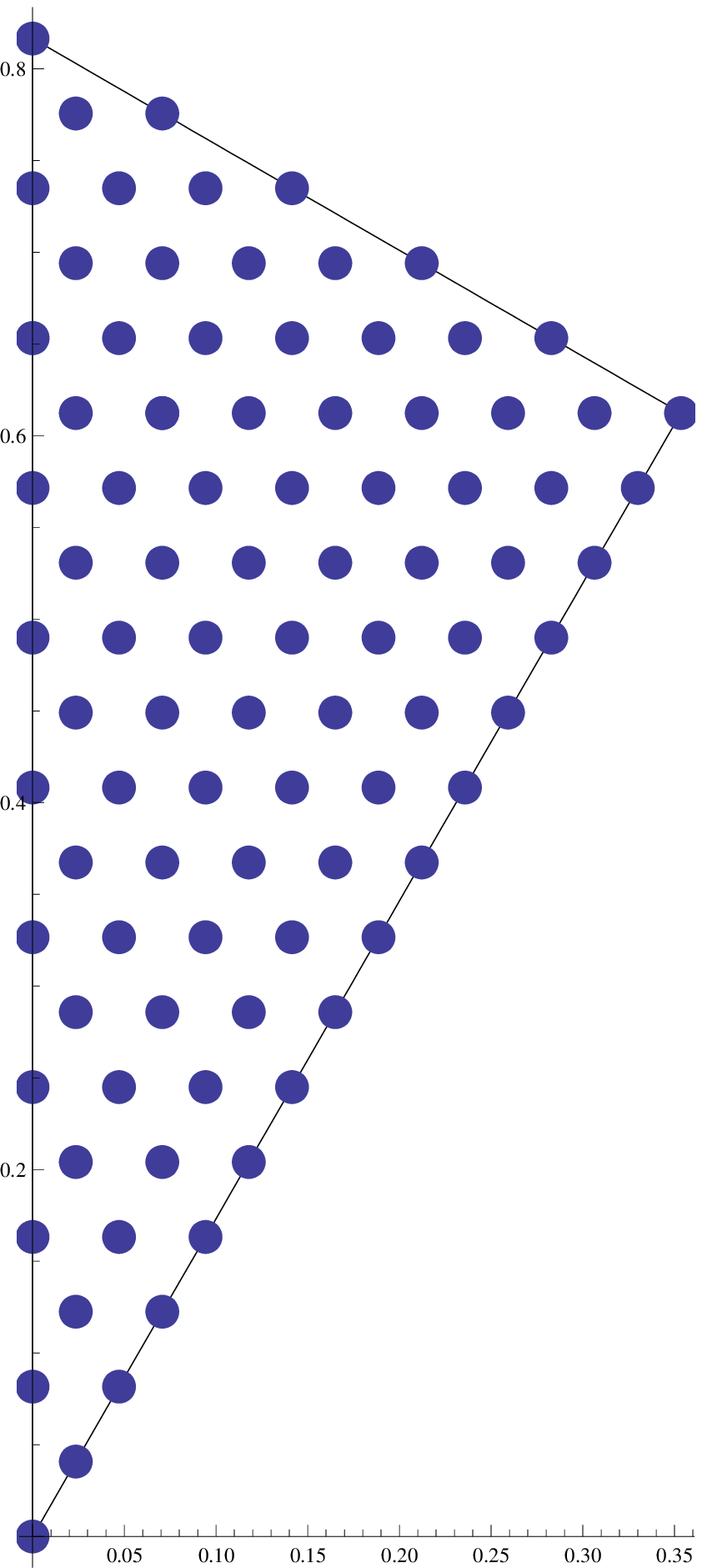} \\
$M=16$ \hskip 3 cm $M=30$
\caption{{\footnotesize The lattice grids $F_3$, $F_6$, $F_{10}$, $F_{16}$ and $F_{30}$.}}
\label{grids}
\end{figure}
Note the relations $F_3\subset F_6\subset F_{30}$ and $F_{10}\subset F_{30}$,  and also
$F_{16} \subset {\hskip -0,3cm /} \,\ F_{30}$.
\end{example}

The size of the grid $|F_{M}(G_2)|$ is known \cite{HP09}. It is the number of points in $F_{M}$.
\begin{gather*}\label{gridsize}
|F_{M}(G_2)|=\left[\tfrac{M}{3}\right]+1+\sum_{i=0}^{\left[\frac M3\right]}\left[\tfrac{M-3i}{2}\right],
\qquad M\in\N\,.
\end{gather*}
where $[\ \cdot\ ]$ denotes the integer part of a number.

\subsection{Fourier analysis on the lattice grids $F_M$}\

In general, discrete orthogonality of $C$ functions of any simple Lie group was first introduced and extensively used in \cite{MP87}. Discrete orthogonality of $G_2$-functions of the $C$ and $S$ families was described in \cite{PZ2,PZ3}. It is included here for completeness. The discrete orthogonality of the $G_2$ functions within the families of $S^L$- and $S^S$-functions is new.

Recall
\begin{equation}\label{C-orthog}
\l C_{(a,b)},C_{(c,d)}\r_M  :=\sum\limits_{s\in F_M} c_s C_{(a,b)}(s)C_{(c,d)}(s)\,,\qquad a,b,c,d\in\Z^{\geq0}\,,
\end{equation}
where the coefficients $c_s=\frac{|W|}{|\operatorname{Stab}_W(s)|}$ count the number of points on the maximal torus that are conjugate to $s\in F$. In the case of $G_2$, the coefficients are given in Table~\ref{coeffs}.

\begin{table}[h]
{\footnotesize
\begin{center}
\begin{tabular}{|c||c|c|c|c|c|c|c|c|}
\hline \rule{0pt}{10pt}
$s\in F_M$&$[*,*,*]$
          &$[*,*,0]$
          &$[*,0,*]$
          &$[0,*,*]$
          &$[*,0,0]$
          &$[0,*,0]$
          &$[0,0,*]$
        \rule{0pt}{10pt}\\[2pt]
\hline \rule{0pt}{10pt}
$c_s$  &$12$
        &$6$
        &$6$
        &$6$
        &$1$
        &$3$
        &$2$
\rule{0pt}{10pt}\\[2pt]
\hline
\end{tabular}\end{center}}
\bigskip
\caption{{\footnotesize The values of coefficients $c_s$ of discrete orthogonality relations \eqref{C-orthog}. Points $c$ are shown in Kac coordinates $[s_0,s_1,s_2]$, where $\star$ denotes any positive integer.}}  \label{coeffs}
\end{table}

More generally, we define the product of functions $f(s)$ and $g(s)$ sampled at the points $s\in F_M$,
\begin{gather}\label{discreteproduct}
\l f,g\r_M:=\sum_{s\in F_M} c_sf(s)\overline{g(s)}\,.
\end{gather}

Combining \eqref{C-orthog} and \eqref{discreteproduct}, we have the formalism for expansion of digital data:
\begin{gather}
f(s)=\sum_\lambda d_\lambda C_\lambda(s)\,,\qquad
    \lambda\in P^+,\quad s\in F_M\,,\label{fourier}\\
d_\lambda=\sum_{s\in F_M} c_sf(s)C_\lambda(s)
   =:\sum_{s\in F_M}M_{\lambda s}f(s)\,.\label{inversefourier}
\end{gather}
The matrix $M=(M_{\lambda s}):=(c_sC_\lambda(s))$ is not dependent on the data function $f(s)$. Therefore, it can be calculated in advance, stored, and used again, thus speeding up the computations.

The range of values of $\lambda$ involved in \eqref{fourier} and \eqref{inversefourier} for each $F_M$ was found\footnote{This fact is the core of the generalization to $G_2$ of the Nyquist-Shannon sampling theorem.} in \cite{HP09}, namely
\begin{gather*}
\lambda \in \Lambda_M := \{(a,b)\in P^+\mid 3a+2b\le M\}.
\end{gather*}

The principal motivation underlying this paper is to generalize the relations \eqref{fourier} and \eqref{inversefourier} to the functions of the families $S$, $S^L$ and $S^S$, thus increasing  digital data processing options.

\subsection{Orthogonality of orbit functions on the lattice grids $F_M$}\

Orthogonality of the orbit functions is what makes the relations  \eqref{fourier} and \eqref{inversefourier} possible. 
\begin{gather}\label{Corthog}
  \l C_{(a,b)},C_{(c,d)}\r_M =  12M^2\cdot \left\{\begin{array}{ll}
\frac{1}{12}, & \textrm{if } a = b = c = d = 0,\\
\frac12, & \textrm{if } a=c=0 \  \mathrm{and} \ 0<b=d<{M \over 2}, \\
& \quad \mathrm{or} \ b=d=0 \ \mathrm{and} \ 0<a=c<{M \over 3},\\
1, & \textrm{if } a=c=0 \ \mathrm{and} \ 0<b=d={M\over 2}, \\
& \quad \mathrm{or} \ 0<a=c, 0<b=d \ \mathrm{and} \ 3a+2b<M,\\
\frac{3}{2}, & \textrm{if } 0<a=c={M \over 3} \ \mathrm{and} \ b=d=0,\\
2, & \textrm{if } 0<a=c,0<b=d \ \mathrm{and} \ 3a+2b=M,\\
0, & \textrm{in other cases}.
  \end{array}\right.
\end{gather}

Orthogonality of $S$-functions on $F_M$ is closely related to the discrete orthogonality of irreducible characters \eqref{character}. It was proven independently in \cite{MP06}.
\begin{gather}\label{Sorthog}
  \l S_{(a,b)},S_{(c,d)}\r_M = 12M^2 \cdot \left\{\begin{array}{ll}
1, & \textrm{if } 0<a=c,0<b=d \ \mathrm{and} \ 3a+2b<M,\\
0, & \textrm{in other cases}.  \end{array}\right.
\end{gather}
The simplicity of \eqref{Sorthog} in comparison with \eqref{Corthog} is due to the fact that $S$-functions are zero on the boundary of $F$. Indeed, replacing $C$-functions in \eqref{C-orthog} by $S$-functions, contributions from the points at the boundary $\partial F$ of $F$ to the sum is eliminated because the $S$-functions vanish at $\partial F$.

The demonstration of the orthogonality of $S^L_\lambda$- and $S^S_\lambda$-functions follows from the argument in \cite{MP06}. One has to take into account at which part of $\partial F$ the functions vanish, and at which part they remain \eqref{atboundaries}.
\begin{gather*}
  \l S^L_{(a,b)},S^L_{(c,d)}\r_M = 12M^2 \cdot \left\{\begin{array}{ll}
\frac12, & \textrm{if } b=d=0 \ \mathrm{and} \ 0<a=c<{M \over 3},\\
1, & \textrm{if } 0<a=c, 0<b=d \ \mathrm{and} \ 3a+2b<M,\\
\frac{3}{2}, & \textrm{if } 0<a=c={M \over 3} \ \mathrm{and} \ b=d=0,\\
2, & \textrm{if } 0<a=c,0<b=d \ \mathrm{and} \ 3a+2b=M,\\
0, & \textrm{in other cases}.
  \end{array}\right.
\end{gather*}

\begin{gather*}
  \l  S^S_{(a,b)},S^S_{(c,d)}\r_M = 12M^2 \cdot \left\{\begin{array}{ll}
\frac12, & \textrm{if } a=c=0 \  \mathrm{and} \ 0<b=d<{M \over 2}, \\
1, & \textrm{if } 0<a=c, 0<b=d \ \mathrm{and} \ 3a+2b<M,\\
0, & \textrm{in other cases}.
  \end{array}\right.
\end{gather*}


\section{Decomposition of products of orbit functions}\label{Decomp}
The product of any pair of orbit functions has definite transformation properties with respect to the Weyl group. It decomposes into the sum of orbit functions with the same transformation properties. Ten different pairs can be formed from the functions of the four families. The product of each pair decomposes into the sum of orbit functions of one family. In Table~\ref{products}, the outcome of the decomposition is shown for the ten pairs.
\begin{table}[h]
{\footnotesize
\begin{center}
\begin{tabular}{|c||c|c|c|c|c|c|c|c|c|c|}
\hline \rule{0pt}{10pt}
product &$S^SS^S$
        &$S^SS$
        &$SS$
        &$S^SC$
        &$S^SS^L$
        &$SC$
        &$SS^L$
        &$CC$
        &$CS^L$
        &$S^LS^L$  \rule{0pt}{10pt}\\[2pt]
\hline \rule{0pt}{10pt}
terms   &$C$
        &$S^L$
        &$C$
        &$S^S$
        &$S$
        &$S$
        &$S^S$
        &$C$
        &$S^L$
        &$C$   \rule{0pt}{10pt}\\[2pt]
\hline
\end{tabular}\end{center}}
\bigskip
\caption{{\footnotesize Structure of the decomposition of the ten types of products of orbit functions. The second row shows which functions appear in all the terms of the decomposition of a product into the sum.}} \label{products}
\end{table}
Decomposition of characters into the sum of irreducible characters is a well known problem with no bounds: the bigger the representations multiplied, the longer the sum of corresponding irreducible characters. In contrast, decomposition of orbit functions is always a finite problem.

In the case of $G_2$, an orbit function is the sum of at most 12 exponential terms. Hence a product of two functions contains at most 144 exponential terms regardless how large their dominant points may be. Therefore at least in principle, it would be possible to write down all of pairs of orbit functions. Given the fact that there are ten types of products, it would not be very practical. Therefore we present here decompositions of only some of the lowest orbit functions and some of the generic ones. The arguments $x$ of each orbit function has been omitted to simplify notation.
\begin{equation}\label{lowdecomp}
\begin{aligned}
C_{(1,0)}C_{(1,0)} &= C_{(2,0)}+2C_{(1,0)}+2C_{(0,3)}+6C_{(0,0)},\\
C_{(0,1)}C_{(0,1)} &= C_{(0,2)}+2C_{(0,1)}+2C_{(1,0)}+6C_{(0,0)},\\
C_{(0,1)}C_{(1,0)} &= C_{(1,1)}+2C_{(0,2)}+2C_{(0,1)}.
\end{aligned}
\end{equation}

Examples of more general products of the orbit functions of $G_2$ and their decompositions:
\begin{align*}
C_{(a,b)}C_{(a,b)}=&12C_{(0,0)}{+}C_{(2a,2b)}{+}2C_{(a, b)}{+}2C_{(a{+}b 0)}
              {+}2C_{(a,0)}{+}2C_{(0,b)}{+}2C_{(2a{+}b,0)}\\
                   &{+}2 C_{(b,3a)}{+}2C_{(0,3a{+}b)}{+}2C_{(0,3a{+}2 b)},\\
C_{(a,0)}C_{(a,0)}=&6C_{(0,0)}{+}C_{(2a,0)}{+}2C_{(a,0)}{+}2C_{(0,3a)},\\
C_{(0,b)}C_{(0,b)}=&6C_{(0,0)}{+}C_{(0,2b)}{+}2C_{(0,b)}{+}2C_{(b,0)},\\
C_{(a,b)}S_{(a,b)}=&{-}2S_{(b,3a)}{+}S_{(2a,2 b)}{+}2S_{(a, b)},\\
C_{(a,b)}S^L_{(a,b)}=&S^L_{(2a,2b)}{+}2S^L_{(2a{+}b,0)}{-}2S^L_{(a,b)}
            {-}2S^L_{(a,0)}{-}2S^L_{(a{+}b,0)},\\
C_{(a,0)}S^L_{(a,0)}=&S^L_{(2a,0)}{-}2S^L_{(a,0)},\\
C_{(a,b)}S^S_{(a,b)}=&S^S_{(2 a, 2 b)}{-} 2 S^S_{(a, b)}{-}2 S^S_{(0, b)}{-}2 S^S_{(0, 3 a{+}b)}{+}2S^S_{(0, 3 a{+}2 b)},\\
C_{(0,b)}S^S_{(0,b)}=&S^S_{(0, 2 b)} {-} 2 S^S_{(0, b)},
\end{align*}\begin{align*}
S_{(a,b)}S_{(a,b)}=&12C_{(0, 0)}{+} C_{(2 a, 2 b)}{+} 2 C_{(a, b)}{-} 2C_{(a{+}b, 0)}{-} 2C_{(a, 0)}{-}2C_{(0, b)}{-}
 2C_{(2 a{+}b, 0)}\\
                   &{+}2 C_{(b, 3 a)}{-} 2C_{(0, 3 a{+}b)}{-} 2C_{(0, 3 a{+}2 b)},\\
S_{(a,b)}S^L_{(a,b)}=&S^S_{(2 a, 2 b)}{-} 2 S^S_{(a, b)}{+} 2S^S_{(0, b)}{+}2 S^S_{(0, 3 a{+}b)}{-}2 S^S_{(0, 3 a{+}2 b)},\\
S_{(a,b)}S^S_{(a,b)}=&S^L_{(2 a, 2 b)}{-} 2 S^L_{(a, b)}{-} 2S^L_{(2 a{+}b, 0)}{+}2 S^L_{(a, 0)}{+} 2S^L_{(a{+}b, 0)},
\end{align*}\begin{align*}
S^L_{(a,b)}S^L_{(a,b)}=&{-}12C_{(0, 0)}{+} C_{(2 a, 2 b)}{+}2 C_{(a, b)}{+} 2C_{(a{+}b, 0)}{+} 2C_{(a, 0)}{-}2C_{(0, b)}{+}
  2C_{(2 a{+}b, 0)}\\
                     &{-} 2 C_{(b, 3 a)}{-} 2C_{(0, 3 a{+}b)}{-}2C_{(0, 3 a{+}2 b)},\\
S^L_{(a,0)}S^L_{(a,0)}=&{-}6 C_{(0, 0)} {+}  C_{(2 a, 0)} {+} 2 C_{(a, 0)} {-} 2 C_{(0, 3 a)},\\
S^L_{(a,b)}S^S_{(a,b)}=&S_{(2 a, 2 b)}{+} 2 S_{(b, 3 a)}{+} 2 S_{(a, b)},\\
S^S_{(a,b)}S^S_{(a,b)}=&{-}12C_{(0, 0)}{+} C_{(2 a, 2 b)}{+} 2 C_{(a, b)}{-}2C_{(a{+}b, 0)}{-}2C_{(a, 0)}{+}2C_{(0, b)}{-}
  2C_{(2 a{+}b, 0)}\\
                       & {-} 2 C_{(b, 3 a)}{+} 2C_{(0, 3 a{+}b)}{+} 2C_{(0, 3 a{+}2 b)},\\
S^S_{(0,b)}S^S_{(0,b)}=&{-}6 C_{(0, 0)} {+}  C_{(0, 2 b)} {+} 2 C_{(0, b)} {-} 2 C_{(b, 0)}.
\end{align*}

\begin{example}\

Curiously, there are precisely 14 points $x\in F$ where all $C$-functions of $G_2$ take integer values. Those points (`rational points') were found in \cite{MPS} and are shown in Table~\ref{efos}. By replacing the $C$-functions in \eqref{lowdecomp} by the corresponding integers, the equalities are maintained. In order to see that, take any of the 14 columns of Table~\ref{efos} and replace the $C$-functions in \eqref{lowdecomp} by the entries in that column from the corresponding rows.

For example, we consider the decomposition of $C_{(1,0)}(x)C_{(1,0)}(x)$ of
\eqref{lowdecomp} at points $x=(0,0)$, $(\tfrac12,0)$, $(\tfrac12,0)$, $(\tfrac12,\tfrac1{12})$. Values of the $C$-functions at these points are taken from the corresponding columns of Table~\ref{efos}.
\begin{alignat*}{5}
&x\in F&
C_{(1,0)}(x)C_{(1,0)}(x)&= C_{(2,0)}(x)
                &&+2C_{(1,0)}(x)
                &&+2C_{(0,3)}(x)
                &&+6C_{(0,0)}(x),\notag\\
&x=(0,0)&
6\times6\qquad   &=\quad6
                &&+\ 2\times6
                &&+\ 2\times6
                &&+\ 6\times1, \notag\\
&x=(\tfrac12,0)&
(-2)\times(-2)\   &=\quad6
                &&+2\times(-2)
                &&+2\times(-2)
                &&+\ 6\times1,\\
&x=(\tfrac13,0)&\quad
(-3)\times(-3)\   &=\ -3
                &&+2\times(-3)
                &&+\ 2\times6
                &&+\ 6\times1, \notag\\
&x=(\tfrac13,\tfrac1{12})&\quad
(-1)\times(-1)\   &=\quad 1
                &&+2\times(-1)
                &&+\ 2\times(-2)
                &&+\ 6\times1. \notag
\end{alignat*}

Analogous equalities hold for all recurrence relations of $G_2$. Unfortunately, the rational points for $S$, $S^L$, and $S^S$ families, where the functions take integer values, are not known. However they do exist because the origin $x=(0,0)$ is clearly one of them.

\end{example}

\section{Arithmetic properties of orbit functions}\label{Arith}

Consider the three functions \eqref{ratios}, the first of which is the Weyl character formula. In this section, we are interested in the cases where the variables $x\in F$ of the functions have rational coordinates in the $\check\omega$-basis. They were extensively studied in \cite{MP84}. The common divisor of the coordinates of such $x$ is the order $M$ of the elements of $G_2$ that belongs to the same conjugacy class of elements represented in $F$ by $x$. Compare a description of such elements
\begin{gather}\label{efo}
\{x\in F\mid x=\tfrac{s_1}M\check\omega_1+\tfrac{s_2}M\check\omega_2\,,\
    M=s_0+2s_1+3s_2,\\
     s_0,s_1,s_2\in\Z^{\geq0},\ \gcd\{s_0,s_1,s_2\}=1\}\notag
\end{gather}
with the lattice points of $F_M$. The difference is that in \eqref{efo} one also requires $\gcd\{s_k\}=1$, i.e., that the Kac coordinates $[s_0,s_1,s_2]$ of $x$ have no common divisor.

Arithmetic properties of the orbit functions are meant to be here the information that can be deduced from the character values of conjugacy classes of elements of finite order in $G_2$.

\begin{figure}[h]
  \centering \includegraphics[angle=90,scale=0.4]{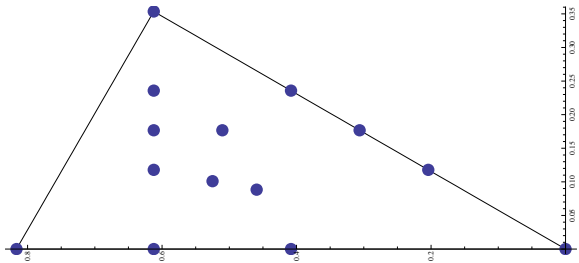}
  \caption{{\footnotesize Rational elements of $G_2$ shown as points in $F$.}}
  \label{ratefos}
\end{figure}

Most interesting are the rational elements, that is the elements of $G_2$ given by the points of $x\in F$ at which the characters $\chi_\lambda(x)$ take integer values for all $\lambda\in P^+$. They are the conjugacy classes of finite symmetries of the root lattice $Q$ of $G_2$. There are precisely 14 conjugacy classes of rational elements in $G_2$. Their position in $F$ is shown in Figure~\ref{ratefos}. They were found in \cite{MPS}. Table~\ref{efos} contains the rational elements and their values at the lowest $C$-functions.

The matrix $(m_{\lambda\mu})$ of the dominant weight multiplicities, see \eqref{character}, is triangular \cite{BMP} with entries 1 on the diagonal. Such a matrix can be inverted so that the $C$-functions are given as a linear combination of characters with integer coefficients. When considering the rational elements at which all the characters are integer valued,  all the $C$-functions also have integer values. Consequently all the functions of \eqref{ratios} have integer values at rational points.

By definition \cite{MP84}, a point $x\in F$ specifies a rational element $g(x)\in G_2$ of some order $M$, i.e., $g(x)^M=1$, provided all powers $g(x)^k$, where $k$ does not divide $M$, are elements conjugate to $g(x)$. When a power $k$ of $g(x)^k$ divides $M$, it is a rational element of lower order, namely $M/k$.


\begin{sidewaystable}
\vskip 10cm
{\footnotesize
\begin{tabular}{|c||c|c|c|c|c|c|c|c|c|c|c|c|c|c|}
\hline \rule{0pt}{10pt}
$M$&$1$&$2$&$3$&$3$&$4$&$4$&$6$&$6$&$6$&$7$&$8$&$8$&$12$&$12$\rule{0pt}{10pt}\\[2pt]
\hline \rule{0pt}{10pt}
$[s_0,s_1,s_2]$&$[1,0,0]$&$[0,1,0]$&$[1,1,0]$&$[0,0,1]$&$[2,1,0]$&$[1,0,1]$&$[4,1,0]$&$[3,0,1]$&$[1,1,1]$&$[2,1,1]$&$[3,1,1]$&$[1,2,1]$&$[3,3,1]$&$[1,4,1]$  \\[2pt]\hline\rule{0pt}{10pt}
$(\tfrac{s_1}M,\tfrac{s_2}M)$&$(0,0)$&$(\frac{1}{2},0)$&$(\frac{1}{3},0)$&$(0,\frac{1}{3})$&$(\frac{1}{4},0)$
&$(0,\frac{1}{4})$&$(\frac{1}{6},0)$&$(0,\frac{1}{6})$&$(\frac{1}{6},\frac{1}{6})$&$(\frac{1}{7},\frac{1}{7})$
&$(\frac{1}{8},\frac{1}{8})$&$(\frac{1}{4},\frac{1}{8})$&$(\frac{1}{4},\frac{1}{12})$&$(\frac{1}{3},\frac{1}{12})$  \\[2pt]\hline\hline\rule{0pt}{10pt}
$C_{(1,0)}$&$6$&$-2$&$-3$&$6$&$-2$&$2$&$1$&$-2$&$1$&$-1$&$-2$&$0$&$-2$&$-1$  \\[2pt]\hline \rule{0pt}{10pt}
$C_{(0,1)}$&$6$&$-2$&$0$&$-3$&$2$&$-2$&$4$&$1$&$-2$&$-1$&$0$&$-2$&$-1$&$-2$  \\[2pt]\hline \rule{0pt}{10pt}
$C_{(1,1)}$&$12$&$-4$&$0$&$-6$&$-4$&$4$&$-4$&$2$&$2$&$5$&$4$&$0$&$2$&$-2$  \\[2pt]\hline\rule{0pt}{10pt}
$C_{(2,0)}$&$6$&$6$&$-3$&$6$&$-2$&$-2$&$-3$&$6$&$-3$&$-1$&$2$&$-2$&$-2$&$1$  \\[2pt]\hline\rule{0pt}{10pt}
$C_{(0,2)}$&$6$&$6$&$0$&$-3$&$-2$&$-2$&$0$&$-3$&$0$&$-1$&$-2$&$2$&$1$&$4$  \\[2pt]\hline\rule{0pt}{10pt}
$C_{(0,3)}$&$6$&$-2$&$6$&$6$&$2$&$-2$&$-2$&$-2$&$-2$&$-1$&$0$&$-2$&$2$&$-2$  \\[2pt]\hline\rule{0pt}{10pt}
$\chi_{(1,0)}$&$14$&$-2$&$-1$&$5$&$2$&$2$&$7$&$1$&$1$&$0$&$0$&$0$&$-1$&$-1$  \\[2pt]\hline\rule{0pt}{10pt}
$\chi_{(0,1)}$&$7$&$-1$&$1$&$-2$&$3$&$-1$&$5$&$2$&$-1$&$0$&$1$&$-1$&$0$&$-1$  \\[2pt]\hline\rule{0pt}{10pt}
$\chi_{(1,1)}$&$64$&$0$&$-2$&$1$&$0$&$0$&$18$&$0$&$0$&$1$&$0$&$0$&$0$&$0$  \\[2pt]\hline\rule{0pt}{10pt}
$\chi^L_{(1,0)}$&$8$&$0$&$-1$&$8$&$0$&$4$&$3$&$0$&$3$&$1$&$0$&$2$&$0$&$1$  \\[2pt]\hline\rule{0pt}{10pt}
$\chi^L_{(0,1)}$&$6$&$-2$&$0$&$-3$&$2$&$-2$&$4$&$1$&$-2$&$-1$&$0$&$-2$&$-1$&$-2$  \\[2pt]\hline\rule{0pt}{10pt}
$\chi^L_{(1,1)}$&$30$&$-2$&$0$&$-15$&$-2$&$-2$&$4$&$1$&$-2$&$2$&$2$&$-2$&$1$&$-2$  \\[2pt]\hline\rule{0pt}{10pt}
$\chi^S_{(1,0)}$&$20$&$-4$&$-1$&$2$&$4$&$0$&$11$&$2$&$-1$&$-1$&$0$&$-2$&$-2$&$-3$  \\[2pt]\hline\rule{0pt}{10pt}
$\chi^S_{(0,1)}$&$8$&$0$&$2$&$-1$&$4$&$0$&$6$&$3$&$0$&$1$&$2$&$0$&$1$&$0$  \\[2pt]\hline\rule{0pt}{10pt}
$\chi^S_{(1,1)}$&$70$&$-2$&$-5$&$-2$&$-2$&$2$&$19$&$-2$&$1$&$0$&$-2$&$0$&$-2$&$-1$  \\[2pt]\hline
\end{tabular}}
\medskip
\caption{{\footnotesize Values of the lowest $C$-functions at all the points in $F$ representing the conjugacy classes of rational elements of $G_2$. The first row contains the order $M$ of an element, the second row shows its Kac coordinates, the 3rd row shows the coordinates of the point in $\check\omega$-basis of $F$. Subsequent rows contain the values of the $C$-functions shown in the head of each row.}}\label{efos}
\end{sidewaystable}

\subsection*{Acknowledgements}\

I would like to express my gratitude to the Doppler Institute of the Czech Technical University for the hospitality extended to me in Prague and in D\v e\v c\'\i n, where the work was started.

I would like to thank Dr. J. Patera for helpful discussions and comments.


\end{document}